\documentclass[12pt]{article}

\newcommand{\bbr}{I\!\! R}
\newcommand{\bbz}{Z\!\!\! Z}

\newcommand{\2}{$^2$}
\newcommand{\3}{$^3$}
\newcommand{\4}{$_4$}

\newcommand{\x}{arXiv:}

\usepackage{graphicx}

\begin{document}
\thispagestyle{empty}
\begin{center}

\null \vskip-1truecm \vskip2truecm {\Large{\bf Inflation, Large
Branes, and the Shape of Space

}}

\vskip1truecm {\large Brett McInnes} \vskip1truecm

 National University of Singapore

email: matmcinn@nus.edu.sg\\

\end{center}
\vskip1truecm \centerline{ABSTRACT} \baselineskip=15pt
\medskip
Linde has recently argued that compact flat or negatively curved
spatial sections should, in many circumstances, be considered
typical in Inflationary cosmologies. We suggest that the ``large
brane instability" of Seiberg and Witten eliminates the negative
candidates in the context of string theory. That leaves the flat,
compact, three-dimensional manifolds --- Conway's
\emph{platycosms}. We show that deep theorems of Schoen, Yau,
Gromov and Lawson imply that, even in this case, Seiberg-Witten
instability can be avoided only with difficulty. Using a specific
cosmological model of the Maldacena-Maoz type, we explain how to
do this, and we also show how the list of platycosmic candidates
can be reduced to three. This leads to an extension of the basic
idea: the conformal compactification of the entire Euclidean
spacetime also has the topology of a flat, compact,
four-dimensional space.

 \vskip3.5truecm
\begin{center}

\end{center}

\newpage

\addtocounter{section}{1}
\section* {\large{1. \emph{Nearly} Flat or \emph{Really} Flat? }}
Linde has recently argued \cite{kn:lindetypical} that, at least in
some circumstances, we should regard cosmological models with flat
or negatively curved \emph{compact} spatial sections as the norm
from an Inflationary point of view. Here we wish to argue that
cosmic holography, in the novel form proposed by Maldacena and
Maoz \cite{kn:maoz}, gives a deep new interpretation of this idea,
and also sharpens it very considerably to exclude the negative
case. This focuses our attention on cosmological models with
\emph{flat, compact} spatial sections.

Current observations \cite{kn:spergel} show that the spatial
sections of our Universe [as defined by observers for whom local
isotropy obtains] are fairly close to being flat: the total
density parameter $\Omega$ satisfies $\Omega$ = 1.02 $\pm$ 0.02 at
95$\%$ confidence level, if we allow the imposition of a
reasonable prior \cite{kn:lahav} on the Hubble parameter. [See
however \cite{kn:padman} for a cautionary note.] The present era
of ``precision cosmology" \cite{kn:primack} is based on the
assumption that the true value of $\Omega$ is even closer to unity
than the observations demand
--- see for example \cite{kn:schechter}. Applications of precision
cosmology depend on this ``almost exactly flat" assumption in a
crucial way: for example, Wang and Tegmark \cite{kn:yun} stress
that without this assumption essentially nothing can be said about
the evolution of the dark energy density. Turning to the
theoretical situation, we find that the leading theory, Inflation
\cite{kn:inflated}\cite{kn:burgess}, also demands values of
$\Omega$ which are extremely close, though not exactly equal, to
unity. Most versions require unity \emph{plus or minus} some small
number [typically \cite{kn:inflated} about 10$^{-4}$].

Of course, Inflation itself explains why the Universe currently
\emph{appears} to be flat: any local evidence of curvature is
``inflated away". But here we wish to propose that this process
merely restores the local spatial geometry to its initial and most
natural global state, namely that of a \emph{perfectly flat,
compact three-dimensional manifold}. That is, we suggest that the
fundamental value of $\Omega$ is \emph{exactly}, not nearly,
unity; this is proposed as an exact initial condition for stringy
cosmology.

The reader is entitled to ask whether the distinction between
approximate and exact initial flatness really has any content. For
it is clear that ordinary, flat $\bbr^3$ can be given a constant
negative curvature of any magnitude, however small, since
hyperbolic space H\3 has this same $\bbr^3$ topology. Similarly,
$\bbr^3$ can be consistently deformed so that it has positive
Ricci curvature everywhere.\footnote{Examples of this can be
constructed, but of course in this case the Ricci curvature cannot
be \emph{constant}, that is, the metric cannot be Einstein, if the
metric is complete. In fact \cite{kn:schoen}, $\bbr^3$ is the only
non-compact three-dimensional manifold which can accept a complete
metric of positive Ricci curvature.} Thus flat $\bbr^3$ can be
deformed in a way which leads to either positive or negative Ricci
curvature, of any magnitude, at every point, and so it is hard to
see how there can be a difference between extremely small
curvature and exactly zero curvature.

This, however, is where the assumption of \emph{compactness} is
crucial. For the topology of an exactly flat compact manifold is
radically different from that of either a positively or a
negatively curved space, whether compact or not. A consequence of
this is that it is impossible to deform a compact flat manifold in
such a way that its sectional curvature is everywhere negative; on
the other hand, it is also impossible to deform it so that even
the \emph{scalar} curvature becomes positive everywhere. [See
\cite{kn:schoenyau} and page 306 of \cite{kn:lawson}.] Of course,
such a space can be locally deformed [by the presence of a galaxy,
say] but not in a way leading consistently to curvature of a
definite sign. If the Universe had spatial sections of this kind,
and if the matter content were smoothed out, then the geometry
would \emph{have} to be exactly flat, as a result of these
extremely deep geometric theorems. Thus, the hypothesis of exact
underlying flatness does make sense if the spatial sections are
compact.

The suggestion that the spatial topology of our Universe is not
trivial is of course an old one \cite{kn:levin}. Some current
interest in this idea focuses on the relation to the AdS/CFT
correspondence
\cite{kn:mcinnesschwarz}\cite{kn:orbifold}\cite{kn:aps}\cite{kn:ross}\cite{kn:ofar}.
Inflation explains why we probably cannot see \cite{kn:cornish}
direct evidence of such non-triviality: the fundamental domain is
inflated to a size larger than the current cosmological horizon.
Nevertheless, the idea that the spatial sections may be compact
continues to attract attention, from many different points of
view. In the specific case of \emph{flat}, compact sections,
discussions include simple models of components of dark energy
\cite{kn:paban}\cite{kn:elizalde} and string/brane gas cosmology
\cite{kn:brand}\cite{kn:brandwat}\cite{kn:watson}\cite{kn:campos}\cite{kn:greene}\cite{kn:danos};
in particular it is interesting that, whether or not string/brane
gas cosmology succeeds in explaining the dimensionality of
observed space, the Brandenberger-Vafa scenario, with its toral
model of \emph{all } spatial directions, is still widely regarded
as a natural initial condition for string cosmology.

Most relevantly for our work here, it has long been known
\cite{kn:zeldovich}\cite{kn:martin} that flat or negatively curved
\emph{compact} spatial sections arise very naturally in quantum
cosmology. More recently, Linde \cite{kn:lindeball} has emphasised
that such constructions are also natural from the Inflationary
point of view; and, more recently still, as we mentioned earlier,
he has strengthened this to the claim that compact but \emph{not
positively curved} spatial sections should be considered to be
typical in Inflationary quantum gravity rather than exotic
\cite{kn:lindetypical}. Linde stresses that there is no conflict,
in Inflationary theory, between the assumption of compactness and
the Inflationary prediction that the effects of compactness should
not be \emph{directly} observable. In fact, the compactness of the
spatial sections may play a vital role in ensuring sufficient
initial homogeneity for Inflation to begin. In this connection, it
has recently been argued \cite{kn:albrecht} that Inflation
\emph{requires} us to take a global viewpoint and not to ignore
the structure beyond the horizon.

It is the objective of this work to argue that the hypothesis of
exact spatial flatness, but \emph{not} negative curvature, is
natural from the \emph{holographic} point of view.

The form of cosmic holography in which we are interested here, due
to Maldacena and Maoz \cite{kn:maoz} is one which adapts the basic
ideas of the AdS/CFT correspondence to the cosmological case. As
in AdS/CFT, the starting point is anti-de Sitter spacetime, but
now transformed into a cosmological spacetime by the introduction
of some kind of matter
\cite{kn:mcinnes}\cite{kn:maoz2}\cite{kn:answering}\cite{kn:maoz3}.
The resulting cosmology has both a Bang and a Crunch, but its
\emph{Euclidean} version is entirely non-singular and has a
well-behaved conformal infinity, on which the dual field theory is
to be defined. Each connected component of this conformal infinity
has \emph{precisely the same topology} as the spatial sections of
the Lorentzian version of the spacetime.\footnote{This picture is
actually consistent with the hypothesis that the spatial sections
are \emph{compact}, for in the generalized Euclidean AdS/CFT
correspondence it is usually desirable for the CFT to be defined
on a compact space; see Section 2.3 of \cite{kn:witten2}.}

If there is a holographic AdS/CFT-style duality here, it follows
that the cosmological model is controlled by a field theory which
does not ``care" how large the spatial sections may be at any
particular time, such as the present. Whatever their size, the
field theory is still sensitive to their structure, including
their topological structure. [For concrete examples of the
profound ways in which non-trivial topology can affect the
behaviour of field theories, see
\cite{kn:dowker1}\cite{kn:dowker2}.] In short, cosmic holography
allows us to probe the global form of the spatial sections,
whether or not\footnote{All current data are \emph{compatible}
with Inflationary expectations regarding spatial curvature and
topology, but this is not to say that alternatives [see for
example \cite{kn:roukema}\cite{kn:opher}\cite{kn:su}] have been
completely ruled out. For the sake of clarity, however, we shall
assume here that they have been.} the fundamental domain is far
larger than the current horizon: it is capable of this because an
AdS/CFT type of duality is a correspondence between the
\emph{entire} bulk and its infinity.

As an application of these ideas, we shall try to constrain the
structure of the spatial sections. We do this with the aid of the
``large brane instability" discussed by Seiberg and Witten
\cite{kn:seiberg}. We shall see that holography rules out negative
curvature for \emph{compact} cosmological spatial sections,
\emph{no matter how small} the curvature may be in magnitude; in
fact, it is possible to make this argument even if the well-known
``BKL" behaviour [describing the growth of anisotropies during the
approach to cosmological singularities] is taken into account. We
shall also see that holography does allow flat, compact spatial
sections, but only if specific conditions are satisfied close to
the singularities.

If the spatial sections of our world are flat and compact, then it
is potentially important to determine which of the ten possible
topologies \cite{kn:wolf} has been selected --- and why. We shall
not completely succeed in fixing the topology, but the list of
candidates will be greatly reduced, from ten to three. One of the
three survivors is the Hantzsche-Wendt space or ``didicosm"
\cite{kn:weeks}\cite{kn:conway}, the most complex of the ten.

We begin with a very brief introduction to a class of cosmological
models \cite{kn:mcinnes}\cite{kn:answering} which generalize those
proposed by Maldacena and Maoz by allowing for a period of
acceleration, in accord with current observations \cite{kn:riess}.
Throughout this discussion, we shall for simplicity ignore all
forms of matter other than the quintessence field; this includes
the inflaton, though we stress that ultimately [as explained in
\cite{kn:lindetypical}] we rely on Inflation to ensure detailed
agreement with current observations. We then explain how these
models are compatible with cosmic holography, laying particular
stress on the stringent conditions imposed by the Seiberg-Witten
instability \cite{kn:seiberg}. Next, we argue that a holographic
one-to-one bulk/infinity correspondence can be maintained only by
extending our basic hypothesis to the entire spacetime: that is,
we propose that the compactification of the [Euclidean] version is
globally conformal to a four-dimensional flat compact manifold. We
will see that this imposes conditions which only a few candidate
topologies [for the three-dimensional sections] are able to meet.
Because we are concerned with the topology [and not with the
precise geometry] of these spaces, it is reasonable to hope that
our conclusions are valid even though our concrete cosmological
model is too simple to be realistic.

Throughout this work we follow Maldacena and Maoz \cite{kn:maoz}
in assuming that the background geometry, prior to the
introduction of some kind of matter, is that of \emph{anti}-de
Sitter spacetime, AdS\4. [See
\cite{kn:lindekalop}\cite{kn:cardenas}\cite{kn:cvetic}\cite{kn:horowitz}\cite{kn:hertog}
for relevant work on AdS-based cosmology.] Of course, many efforts
have been made to develop cosmic holography on a de Sitter-like
background; see \cite{kn:klemm} for a very clear analysis of the
current state of such attempts. Constraints on cosmic topology can
also be developed in that context: see
\cite{kn:mcinnesschwarz}\cite{kn:orbifold}\cite{kn:aps}.

\addtocounter{section}{1}
\section*{\large{2. Flatness, Acceleration, and Breitenlohner-Freedman}}
For our subsequent discussions it will be very helpful to have a
concrete model of the various physical mechanisms to be
considered. In this section, we introduce an extremely simple
cosmological model which can play this role. No claim is made that
this model itself is realistic, though possibly it could be made
so by superimposing matter, radiation, inflaton and other fields
on the simple spacetime to be defined below.

The basic cosmological model we shall consider is one with a Bang
and a Crunch. There are in fact very general arguments
\cite{kn:giddings} which suggest that the ultimate state of our
Universe will be a Crunch of the kind that arises naturally when
potentials are allowed to be negative \cite{kn:felder}. If our
Universe is \emph{now} anti-de Sitter-like --- something that is
not excluded by observations, since such spacetimes can
accelerate, though only temporarily \cite{kn:mcinnes} --- then
this is a straighforward consequence of having a negative
cosmological constant in the background. But even if the present
state of the Universe is de Sitter-like, this probably corresponds
to a metastable state which eventually fluctuates or tunnels to an
\emph{anti}-de Sitter-like basin of attraction of some potential.
[The alternative is decompactification, but this possibility only
arises if one has some argument which rules out negative
potentials.]

Maldacena and Maoz \cite{kn:maoz} analyse Bang/Crunch spacetimes
with metrics of the form
\begin{equation}\label{eq:DRONGO}
g^-_{\mathrm{MM}} = -\; (dt^-)^2\; +\; a^-(t^-)^2\;g^+(\Sigma),
\end{equation}
where t$^-$ is proper time, $a^-(t^-)$ is the scale factor [which
vanishes at both ends of some finite interval], and $g^+(\Sigma)$
is a metric on a \emph{time-independent} three-dimensional
Riemannian manifold $\Sigma$ which acts as a model for the spatial
sections. [Throughout this work, we use a + superscript to
indicate a Euclidean coordinate or field, a negative sign for its
Lorentzian counterpart.] Note that such metrics do not take into
account the evolution of anisotropies, which we shall consider, in
specific cases, in later sections of this work.

One way of obtaining spacetimes of this kind is to introduce
matter into anti-de Sitter spacetime, allowing it to act on the
geometry in accord with Einstein's equation. The result is
typically a Bang/Crunch spacetime. Maldacena and Maoz observe that
the Euclidean version will in general be non-singular and
asymptotically hyperbolic [that is, asymptotically like Euclidean
AdS]. It will therefore have a well-defined conformal boundary.
The hope is that, in some way that is not yet fully understood,
the non-singular Euclidean boundary ``replaces" the singularities
of the Lorentzian version. A field theory on the boundary should
give a holographic description of the bulk in the familiar
way.\footnote{This is the sense in which we shall understand
``holography". Note that other interpretations, involving entropy
bounds, may not be consistent with Maldacena-Maoz cosmologies: see
in particular \cite{kn:lindekalop}.} Notice that, by contrast, de
Sitter spacetime does not have a holographic Euclidean version,
since the usual Euclidean version of dS\4, the four-sphere, has no
boundary. In this sense, Euclidean holography favours AdS\4 over
dS\4 as the fundamental ``background" for cosmology.

Maldacena and Maoz also observe that the Euclidean versions of
their Bang/Crunch cosmologies are topologically non-trivial: they
refer to such spaces as Euclidean ``wormholes''. For this reason
they use particular matter configurations such as Yang-Mills
merons and instantons to construct their cosmological models.
Unfortunately, such matter cannot lead to cosmic acceleration. On
the other hand, ordinary scalar matter, in the guise of
``quintessence", can easily lead to acceleration, but it cannot
generate a topologically non-trivial Euclidean ``spacetime"
\cite{kn:wald}\cite{kn:coule}. We are thus led, as in
\cite{kn:GS}, to consider a \emph{Euclidean axion} as the matter
content of the Euclidean version of the spacetime; for axions
appear to be unique in leading \emph{both} to acceleration
\emph{and} to topological non-triviality. Other forms of matter
and radiation, as well as the inflaton, have well-known effects on
the expansion history, so for simplicity we shall not consider
them here.

Motivated by the discussion of quintessence superpotentials in
\cite{kn:susskind}, in \cite{kn:answering} we proposed to develop
a Euclidean axion cosmology by postulating a superpotential. Since
the potential should be periodic for an axion, the same applies to
the superpotential; and since the axion field $\varphi^+$ is a
pseudo-scalar, it is natural to restrict attention to
superpotentials which are odd in $\varphi^+$. Thus we consider
superpotentials of the form
\begin{equation}\label{eq:ABRACADABRA2}
\textup{W}^+(\varphi^+)\; =\; \sum_{k=1}^\infty\;\textup{C}_k \,
\textup{sin}(\mathrm{k}\,\sqrt{{{4\pi}\over{\varpi}}}\;\;\varphi^+),
\end{equation}
where $\varpi$ is a  positive constant. If we take only one term
for simplicity, we can assume it to be the first; requiring the
potential to yield the usual negative cosmological constant for
pure Euclidean AdS\4, with all sectional curvatures equal to
$-$1/L\2, when (W$^+$)$^\prime$ vanishes, we can fix the constant
C, and so we obtain
\begin{equation}\label{eq:C}
\textup{W}^+(\varphi^+)\; =\; {{1}\over{16\pi \textup{L}}}\;
\textup{sin}(\sqrt{{{4\pi}\over{\varpi}}}\;\;\varphi^+).
\end{equation}
Higher-order terms in the original expansion
(\ref{eq:ABRACADABRA2}), which we shall consider later, are
obtained by replacing $\varpi$ by $\varpi$/k$^2$.

The potential corresponding to W$^+$ may be written as
\begin{equation}\label{eq:D}
\textup{V}^+(\varphi^+)\; =\; - {{3}\over{8\pi
\textup{L}^2}}\;+\;\textup{V}^+_{\textup{Axion}},
\end{equation}
where
\begin{equation}\label{eq:E}
\textup{V}^+_{\textup{Axion}}\;=\;{{3 \;-\;\varpi^{-1}}\over{8\pi
\textup{L}^2}}\;\textup{cos}^2(\sqrt{{{4\pi}\over{\varpi}}}\;\;\varphi^+).
\end{equation}
Thus we are effectively considering Euclidean AdS\4, with
``energy" density $-$3/(8$\pi$L$^2$), into which we have
introduced a matter field with a potential V$^+_{\textup{Axion}}$.

In accordance with our hypothesis that the spatial sections of our
cosmological model are to be flat and compact, we recall
\cite{kn:wolf} that every compact flat three-manifold can be
expressed as T$^3$/F, where T$^3$ is the three-torus and F is a
small finite group [which is in fact isomorphic to the holonomy
group of the manifold]. Locally, therefore, we can use the usual
angular coordinates on a three-torus [taken to be cubic for
convenience], and the Euclidean metric will have the general form
\begin{equation}\label{eq:I}
g^+\; = \;(dt^+)^2 \;+\; A^2\;a^+(t^+)^2[d\theta_1^2\; +
\;d\theta_2^2 \;+\; d\theta_3^2],
\end{equation}
where A measures the circumferences of the torus when
$a^+$(t$^+$), the Euclidean scale factor [which we can abbreviate
to $a^+$], is equal to unity.

The solution for $a^+$ in the present case, obtained by solving
\cite{kn:mcinnes} the Einstein equations with the potential
(\ref{eq:E}) [and a canonical kinetic term] superimposed on
Euclidean AdS\4, yields the  metric
\begin{equation}\label{eq:M}
g^+(\varpi,A) = (dt^+)^2 +
A^2\;\textup{cosh}^{2\varpi}({{t^+}\over {\varpi
\textup{L}}})\;[d\theta_1^2 + d\theta_2^2 + d\theta_3^2].
\end{equation}
This is entirely non-singular. If we embed this space as the
interior of a manifold-with-boundary, then the boundary has two
connected components at t$^+$ = $\pm\infty$. However, it was
argued in \cite{kn:answering} that holography dictates that these
two components should be topologically identified, and that is
what we propose to investigate below.

The Lorentzian version of all this is significantly different: we
now have a Bang/Crunch spacetime with metric
\begin{equation}\label{eq:L}
g^-(\varpi,A) = -\;(dt^-)^2 +
A^2\;\textup{cos}^{2\varpi}({{t^-}\over {\varpi
\textup{L}}})\;[d\theta_1^2 + d\theta_2^2 + d\theta_3^2].
\end{equation}
Contrary to what is often said, such cosmologies can be perfectly
compatible with current observations, a point stressed recently by
Wang et al \cite{kn:kratochvil}. Notice that this metric allows
for a period of accelerated expansion provided that $\varpi$ is
not too small; in fact there is such an interval if $\varpi$ $>$
1. The Lorentzian version of $\varphi^+$, denoted $\varphi^-$, is
a quintessence field
\cite{kn:ratra}\cite{kn:carroll}\cite{kn:nesseris} with an
exponential-like potential given by
\begin{equation}\label{eq:H}
\textup{V}^-_{\textup{Quintessence}}\;=\;{{3
\;-\;\varpi^{-1}}\over{8\pi
\textup{L}^2}}\;\textup{cosh}^2(\sqrt{{{4\pi}\over{\varpi}}}\;\;\varphi^-);
\end{equation}
this is superimposed on an AdS\4 geometry with cosmological
constant $-$3/L$^2$. The energy density of $\varphi^-$ can be
computed in terms of the Lorentzian scale function $a^-$(t$^-)$
[which we abbreviate to $a^-$]; the result \cite{kn:mcinnes} is
\begin{equation}\label{eq:EXTRA}
\rho\,(\varphi^-)\;=\;{{3}\over{8\pi\textup{L}^2}}\;(a^-)^{-2/\varpi}.
\end{equation}
The total energy density is the sum of this and the energy density
of the background AdS$_4$. If we had taken the k-th order term in
(\ref{eq:ABRACADABRA2}) instead of the first, then the density of
$\varphi^-$ would vary as $(a^-)^{-2\,k^2/\varpi}$; so the k = 1
term dominates when the Universe is large, while the higher order
terms in the Fourier expansion are important very near to the Bang
and the Crunch.

Clearly the Lorentzian metric $g^-(\varpi,A)$ given by
(\ref{eq:L}) is not asymptotically AdS. Nevertheless its Euclidean
version, given by (\ref{eq:M}), \emph{is} asymptotically
hyperbolic, that is, asymptotically similar to Euclidean AdS.
Since the Maldacena-Maoz formulation of cosmic holography is based
on an interplay between the Euclidean and Lorentzian versions, any
constraint on the parameters which we can derive from this fact
must be accepted as physically relevant. A fundamental example of
such a constraint is the Breitenlohner-Freedman bound
\cite{kn:freed}, which, as explained in \cite{kn:witten2}, is also
valid in the Euclidean case. [Notice that the Euclidean space is
compactified only in some directions: its volume is infinite
towards either component of the boundary. The discussion in
\cite{kn:witten2} applies here.] The BF bound imposes a very
interesting condition on $\varpi$, as we now explain.

The field $\varphi^+$ does not decay to zero towards either t$^+$
= $\infty$ or $-\infty$, but rather to
$\pm{{\pi}\over{2}}\sqrt{\varpi/4\pi}$, respectively: this can be
seen from the explicit solution for it,
\begin{equation}\label{eq:K}
\varphi^+\;=\;\pm\sqrt{{{\varpi}\over{4\pi}}}\;\textup{cos}^{-1}(\textup{sech}({{t^+}\over{\varpi\textup{L}}})).
\end{equation}
This behaviour is necessary in order to ensure that the total
energy density should tend to the AdS\4 value $-$3/8$\pi$L$^2$
near infinity [see equation (\ref{eq:E})]. Concentrating on the t
$\rightarrow$ $+\infty$ end of the manifold, we therefore find it
convenient to define a new field $\psi^+$ by
\begin{equation}\label{eq:alphadalpha}
\psi^+\;=\;{{\pi}\over{2}}\sqrt{\varpi/4\pi}\;-\;\varphi^+.
\end{equation}
Substituting this into equation (\ref{eq:E}) we see that the mass
of $\psi^+$ is given by
\begin{equation}\label{eq:alpha}
\mathrm{m}^2\;=\;{{3 \;-\;\varpi^{-1}}\over{\varpi \textup{L}^2}}.
\end{equation}
In general one can expect an AdS/CFT-style correspondence to break
down \cite{kn:gubser} if the Breitenlohner-Freedman bound fails;
since we are of course ultimately interested in establishing a
correspondence of this kind for cosmology, we must ensure that the
BF bound is satisfied here. In four dimensions this bound is m$^2$
$\geq$ (3/4)$\Lambda$, where $\Lambda$ is the negative
cosmological constant of an AdS background. That is, m$^2$ can be
negative without causing any instability, as long as it is not too
negative. Here this bound becomes
\begin{equation}\label{eq:betababy}
{{3 \;-\;\varpi^{-1}}\over{\varpi
\textup{L}^2}}\;\geq\;{{-\;9}\over{4\mathrm{L}^2}},
\end{equation}
whence we have for positive $\varpi$
\begin{equation}\label{eq:beta}
\varpi\;\geq\;{{2}\over{1\;+\;\sqrt{2}}}\;\times {{1}\over{3}}\;.
\end{equation}
Thus the parameter $\varpi$ is allowed to go below the value 1/3,
which means that V$_{\mathrm{Axion}}^+$ [equation (\ref{eq:E})] is
allowed to be negative. However, the lowest value of $\varpi$
allowed by (\ref{eq:beta}) is not very far below 1/3; it is in
fact equal to about 82.8\% of 1/3.

In fact, cosmological data \cite{kn:answering} require the basic
value of $\varpi$ to be quite large; in particular, there is a
period of cosmic acceleration, as observed, if and only if
$\varpi$ is greater than unity. However, our discussion here is
based on the assumption that we take only the \emph{first} term in
the expansion (\ref{eq:ABRACADABRA2}). If we drop this assumption,
then (\ref{eq:beta}) can be interpreted as requiring us to
truncate the expansion in such a way that if k labels the final
term, then not just $\varpi$ but also $\varpi$/k$^2$ satisfies the
inequality. In view of the discussion around equation
(\ref{eq:EXTRA}), this last term will be the dominant one near to
the Bang and the Crunch in the Lorentzian version of the
spacetime.

Combining all these results, we conclude that our Euclidean axion
is governed by a superpotential given by a finite sum in equation
(\ref{eq:ABRACADABRA2}). We can ignore all terms in the sum apart
from the first [which dominates when the Universe is large] and
the last [which dominates near to the Bang and the Crunch]. The
quintessence density will grow very rapidly near to the
Bang/Crunch: it can in fact grow [as $a^-$ tends to zero] more
rapidly than $(a^-)^{-\;6}$. However, evaluating the right side of
(\ref{eq:beta}), we find that the maximum rate at which the
density can tend to zero is as [approximately] $(a^-)^{-\;
7.2426}$. This ``window" between the number 6 and a value just
over 7.2 will be considered in detail below.

Having introduced a concrete example of a holographic cosmology,
we can turn to the question of how holography influences the
structure of the spatial sections of spacetimes of this general
kind.

\addtocounter{section}{1}
\section*{\large{3. Flatness, Holography and Seiberg-Witten Instability}}
Linde \cite{kn:lindetypical} argues that compact spatial sections
are favoured by Inflationary theory. There are in fact several
strong advantages in compact sections: for example, because
compact sections are [under some circumstances] circumnavigable,
it is easy and natural in such cosmologies to arrange for
sufficient homogenization for Inflation to begin. On the other
hand, \emph{positive} curvature is generically disfavoured in
quantum-gravitational studies of initial conditions for Inflation
\cite{kn:martin}. Thus Inflationary quantum gravity firmly directs
our attention towards either flat or negatively curved compact
spatial sections. There is of course an enormous number of such
manifolds, but we shall now see that this number is drastically
reduced when we study Bang/Crunch cosmologies from the AdS/CFT
point of view.

In \cite{kn:seiberg}, Seiberg and Witten have studied the
extension of the AdS/CFT correspondence to general geometries of
the AdS type: that is, they considered the consequences of doing
string theory on non-compact Euclidean spaces, with negative Ricci
curvature, admitting a conformal compactification in the sense of
Penrose. One of their more remarkable findings was that BPS branes
``near" to the conformal boundary [``large branes"] will give rise
to an instability if the conformal structure at infinity is
represented by a metric of negative \emph{scalar} curvature. [When
discussing compact conformal manifolds, we can, without loss of
generality, assume that the scalar curvature is a constant of
arbitrary magnitude but of a fixed sign \cite{kn:schoenyam}.] The
unexpected role of the scalar curvature is a strong hint that this
instability is ``holographic", for one knows that the scalar
curvature is an essential component of the conformally invariant
Laplace operator,
\begin{equation}\label{eq:COFFEE}
\Delta_{\mathrm{CONFORMAL}}\;=\;\Delta\;+\;{{\mathrm{n}\;-\;2}\over{4(\mathrm{n}\;-\;1)}}\;\mathrm{R},
\end{equation}
defined by the conformal structure at [n-dimensional] infinity.
[It is important to note that everything we say here is based on
the assumption that n is \emph{greater than 2}. The case of
two-dimensional boundaries is special and will not be considered
here.] Indeed, Seiberg and Witten were able to show that negative
scalar curvature does induce the instability in the field theory
at infinity that holography demands given the large brane
instability in the bulk. Seiberg-Witten instability has been
subjected to a deep study recently in \cite{kn:buchel} and
\cite{kn:porrati}; it represents a fundamental constraint on
possible boundary geometries and topologies in any generalized
version of the AdS/CFT correspondence. For it is clear that it
would not be consistent to ignore the effects of such unstable
processes on the underlying geometry, and these effects could be
drastic.

This comment applies with particular force in the context of
Maldacena-Maoz holography \cite{kn:maoz}. For here the idea is
that, however singular the Lorentzian cosmology may be, its
Euclidean version should be sufficiently well-behaved that there
are asymptotically AdS regions which are not, for example, cut off
by some kind of disturbance resulting from the unrestrained growth
of large branes in those regions. Thus Seiberg-Witten instability
\emph{must} be avoided in cosmic holography.

The relevance of all this arises from the following simple
observation: the spatial sections of the particular spacetimes
considered in the previous section, and by Maldacena and Maoz,
have the \emph{same} conformal geometry as the space on which the
dual theory is defined; for example, it is clear that if the
manifold with metric given by equation (\ref{eq:M}) is embedded as
the interior of a manifold-with-boundary, then each component of
the boundary has the structure of the flat space T\3/F, with its
``flat" conformal structure. An analogous statement would hold if
we considered a similar spacetime but with negatively curved
spatial sections. If the Maldacena-Maoz cosmologies are a correct
implementation of string theory in cosmology, it therefore follows
that string theory \emph{predicts that the spatial sections of our
Universe cannot be negatively curved}; indeed, they cannot even
have negative scalar curvature. However, this argument ignores
perturbations. We will deal with these after introducing some
mathematical machinery.

The first result we need is the \emph{Kazdan-Warner
classification} \cite{kn:kazdan} --- see \cite{kn:nardmann} for a
recent discussion --- of all compact manifolds of dimension at
least three. This is concerned with the following question: given
such a manifold and any smooth function S on it, does there exist
a metric on that manifold having S as its scalar curvature? This
is ultimately a question about the ``\emph{deformability}" of the
manifold.\footnote{It is interesting that \emph{Lorentzian}
compact manifolds are probably \cite{kn:nardmann} arbitrarily
``deformable" in this sense.} For example, can a sphere [of
dimension greater than two] be deformed to such an extent that its
scalar curvature becomes negative \emph{everywhere}? Such
questions are answered by the Kazdan-Warner classification
theorem:

\bigskip
\noindent THEOREM [Kazdan-Warner]: \emph{All compact manifolds of
dimension at least three fall into precisely one of the following
three classes:}

\medskip
\noindent[P] \emph{On these manifolds, every smooth function is
the scalar curvature of some Riemannian metric.}

\medskip
\noindent[Z] \emph{On these manifolds, a smooth function can be a
scalar curvature of some Riemannian metric if and only if it
either takes a negative value somewhere, or is identically zero.}

\medskip
\noindent[N] \emph{On these manifolds, a smooth function can be a
scalar curvature of some Riemannian metric if and only if it takes
a negative value somewhere.}

\bigskip

For example, spheres are evidently not in [Z] or [N], so they must
be in [P]. [It follows that a sphere of dimension at least three
\emph{can} be deformed in such a way that its scalar curvature is
negative everywhere --- see \cite{kn:porrati} for an explicit
construction.] It can be shown [using some deep theorems to be
discussed below] that compact manifolds of negative sectional
curvature are in [N]. This means that \emph{every} conformal
structure on such a manifold is represented by a metric of
constant negative scalar curvature: \emph{no matter how we deform
it, its scalar curvature can never vanish or become positive
everywhere}. Thus, the Seiberg-Witten instability in this case is
particularly radical, since it is independent of the choice of
metric and must arise from the topology of the space
--- the Kazdan-Warner classification depends only on the
[differential]\footnote{By this we mean that, in some examples of
high-dimensional topological spaces which can admit more than one
differentiable structure, the KW class can change if the
differentiable structure is changed, even if the underlying
topological structure does not change. But this cannot happen in
the cases considered in this work.} topology of the manifold. One
says that the instability is \emph{induced topologically}
\cite{kn:induced}.

This topological aspect of Seiberg-Witten instability has a direct
physical consequence, as follows. The classical
Belinsky-Khalatnikov-Lifschitz analysis of the approach to
cosmological singularities [see \cite{kn:turok1} for a recent
discussion] would lead one to expect that, as a Bang or a Crunch
is approached, the geometry of the spatial sections would become
more and more anisotropic, and this distortion might well become
so extreme that the precise nature of the conformal structure
induced at Euclidean infinity would no longer be clear. Now,
however, we see that such anisotropies are irrelevant: no matter
how complicated they may be, the scalar curvature induced at
Euclidean infinity can never be positive or zero --- no amount of
distortion can avert Seiberg-Witten instability in this case. For
whatever happens to the conformal geometry during the evolution,
the topology of the spatial sections does not change, and the
topology of conformal infinity remains that of a space on which
\emph{every} metric defines a conformal structure with negative
scalar curvature. We conclude that \emph{holography totally
forbids spatial sections of negative curvature}, even if
perturbations are taken into account.

Notice that the theory forbids negative curvature of any
magnitude, no matter how small, because in any case it does not
make sense to speak of ``small" curvatures on the boundary [which
only has a conformal structure, not a Riemannian metric]. Thus
there is indeed a real distinction between \emph{extremely small}
negative curvature and \emph{zero} curvature on the bulk spatial
sections [which do of course have a Riemannian structure]. This
distinction is a direct reflection of the holographic nature of
Maldacena-Maoz cosmology.

To summarize, we have here a very strong prediction from cosmic
holography: the theory could not be saved if any value of $\Omega$
below unity were confirmed by observation. It is interesting to
note that, until the discovery of cosmic acceleration, the
cosmological data actually pointed strongly towards negative
spatial curvature; so we have an example in which cosmic
holography makes a statement which might easily have been
falsified.

Now let us turn to the case of principal interest to us:
cosmological models with flat, compact boundaries and spatial
sections. Seiberg and Witten did not consider the case where the
scalar curvature of the boundary is zero. Here the analysis
depends on higher-order terms \cite{kn:porrati} in the expansion
of the brane action, and unfortunately it is difficult to give a
general statement of the precise conditions needed to avert
instability. However, much can be learned regarding this case by
studying ground states for AdS black holes with flat, compact
event horizons; for these spacetimes have flat conformal
structures on conformal infinity. The ground state for such black
holes is not anti-de Sitter spacetime but rather the ``AdS
instanton" with [Euclidean] metric \cite{kn:HM} given in (n+1)
dimensions by

\begin{equation}\label{eq:HM}
g^+(\mathrm{AdSI}) \;=\;
{{\mathrm{L}^2}\over{\mathrm{r}^2}}\,(1\;-\;{{\mathrm{r_0}^{\mathrm{n}}}\over{\mathrm{r}^{\mathrm{n}}}})^{-1}\,
\mathrm{dr}^2\;+\;
({{\mathrm{r}^2}\over{\mathrm{L}^2}})\,[(\mathrm{dt}^+)^2
\;+\;(1\;-\;{{\mathrm{r_0}^{\mathrm{n}}}\over{\mathrm{r}^{\mathrm{n}}}})\,
\mathrm{dx}^2\;+\;\sum_1^{n-2}\,(\mathrm{dx}^i)^2] .
\end{equation}
Here x and x$^i$ are coordinates on the circumferences of circles
of various radii; that is, they are proportional to angles. In the
Euclidean case, the ``time" coordinate too is angular. The
conformal structure at infinity [r $\rightarrow\;\infty$] is
represented by the flat metric
\begin{equation}\label{eq:HMM}
g^+(\mathrm{AdSI},\;\infty) \;=\; (\mathrm{dt}^+)^2 \;+\;
\mathrm{dx}^2\;+\;\sum_1^{n-2}\,(\mathrm{dx}^i)^2,
\end{equation}
and this is a metric on a \emph{compact} manifold, since all of
the coordinates are angular. Thus the structure at infinity for
the AdS instanton is precisely a compact, flat, n-dimensional
manifold. The very fact that the instanton \emph{is} a
well-behaved, unique ground state
\cite{kn:surya}\cite{kn:gall1}\cite{kn:gall2} for these black
holes strongly suggests that vanishing scalar curvature on the
boundary is compatible with a stable field theory there, dual to
one of these physically well-defined bulk configurations. Thus, we
do have a large class of examples in which zero scalar curvature
at the boundary is not pathological. While there undoubtedly exist
other examples in which it is, one expects that these examples
must involve highly intricate geometric constructions, not the
very simple structures we are considering here.

For concreteness, and in order to avoid giving an analysis which
is too model-dependent, we shall assume that scalar-flat
boundaries of Maldacena-Maoz cosmologies --- which are after all
geometrically much simpler than AdS black holes with flat event
horizons
--- do not lead to large brane instabilities in the bulk. Under
this assumption, the cosmological model we considered above is of
course stable in the Seiberg-Witten sense, since it is clear that
the conformal structure induced on both connected components of
Euclidean infinity is represented by a flat, hence scalar-flat,
metric. As in the negatively curved case, however, one has to
consider whether perturbations can disturb this simple picture.
For a flat manifold can be deformed: a generic distortion produces
a new conformal structure not represented by a flat metric. To
assess the consequences of this, we need some further results in
global differential geometry.

First, we need the concept of an \emph{enlargeable} manifold
[\cite{kn:lawson}, page 302]. These are n-dimensional manifolds M
such that, given any positive $\epsilon$, there exists an
orientable Riemannian covering M$^*$ and a map f [which is
constant at infinity and of non-zero degree] from M$^*$ to the
Riemannian n-sphere of curvature unity, where f contracts all
lengths by a factor of at least $\epsilon$. In other words, M must
have ``arbitrarily large" covering spaces. Notice that
enlargeability is a topological condition. Clearly all compact
flat manifolds are enlargeable.

The work of Schoen, Yau \cite{kn:schoenyau}, Gromov, and Lawson
[\cite{kn:lawson}, page 306] can be summarized as follows:

\bigskip
\noindent THEOREM [Schoen-Yau-Gromov-Lawson]: \emph{There is no
metric of positive scalar curvature on any compact enlargeable
spin manifold.}
\bigskip

It follows that compact enlargeable spin manifolds can never be in
Kazdan-Warner class [P]. Now tori are compact, enlargeable, and
spin; hence, no matter how a torus is deformed, the scalar
curvature can never become positive everywhere, and it follows
that the same is true of any quotient of a torus. Since every flat
compact manifold is a quotient of a torus, we see that this
statement is true of any compact flat manifold. On the other hand,
it is obvious that flat compact manifolds are not in Kazdan-Warner
class [N]. It follows that they are in [Z]. But this means that
the only way to avoid a negative scalar curvature metric on these
spaces is to ensure that the scalar curvature is precisely zero
everywhere. This appears to be a strong constraint. In fact, it is
\emph{far stronger} than it seems. For Gromov and Lawson,
extending a theorem of Bourguignon, were able to prove
[\cite{kn:lawson}, page 308] the following result.

\bigskip
\noindent THEOREM [Bourguignon-Gromov-Lawson]: \emph{If a metric
on a compact enlargeable spin manifold has zero scalar curvature,
then that metric must be exactly flat, that is, the curvature
tensor must vanish everywhere.}
\bigskip

\noindent This is a remarkable result: the vanishing of a single
scalar invariant, the scalar curvature, forces the \emph{entire}
curvature tensor to vanish exactly on these manifolds. Recall now
Schoen's theorem \cite{kn:schoenyam} to the effect that any
conformal structure on a compact manifold is represented by a
metric with constant scalar curvature; recall also that a smooth
function on a manifold in KW class [Z] has to be negative
somewhere if it is the scalar curvature of some metric, unless it
is exactly zero. Combining all these observations, we have the
following statement:

\bigskip
\noindent COROLLARY: \emph{Let} g \emph{be a metric on a manifold
with the topology of a compact flat manifold. Then unless} g
\emph{itself is conformal to a flat metric, it is conformal to a
metric of constant negative scalar curvature.}
\bigskip

\noindent That is, if such a manifold is a component of the
conformal boundary of a manifold of the kind considered by Seiberg
and Witten, and if a flat metric on the boundary is distorted,
however slightly, so that it ceases to be conformally flat, then
the system will become unstable to the production of large branes.
The situation here regarding Seiberg-Witten instability is thus
almost as severe as it is in the negatively curved case: the
instability can be avoided only if the boundary is perfectly
[conformally] flat.

These deep geometric results thus impose an extremely demanding
self-consistency check on our proposal. For the conformal
structure at Euclidean infinity is obtained by taking a suitable
\emph{limit} of the metric on the spatial sections, after removing
the conformal factor. [See the following section for the details.]
This means that, on the Lorentzian side, we have to ensure that
the spatial sections tend to become increasingly flat [again after
removing the conformal factor] as both the Bang and the Crunch are
approached in cosmologies like the one discussed in the previous
section, with the Lorentzian metric given by equation
(\ref{eq:L}). That is of course trivial for this precise metric,
but this simplicity is based on the assumption that no other form
of matter is present. If we introduce small anisotropies
corresponding to local concentrations of matter or radiation, it
is far from clear that the spatial sections will be so
well-behaved near to the Bang and to the Crunch. Indeed, the
Belinsky-Khalatnikov-Lifschitz analysis mentioned above indicates
that under small perturbations a generic spacetime with ordinary
matter sources can be expected to develop severe anisotropies as
one approaches a Bang or a Crunch, and so one would \emph{not} in
general expect a more realistic version of (\ref{eq:L}) to induce
flat metrics on the spatial sections at very early or very late
times; therefore it is far from clear that the conformal structure
at infinity will be represented by a perfectly flat metric.

We shall now see how this problem is naturally avoided by the
cosmological models introduced in the preceding section, for some
values of the fundamental parameter $\varpi$ but not for others.

\addtocounter{section}{1}
\section*{\large{4. Ensuring Flatness at Infinity}}
In order to discuss anisotropies, we need to recall some aspects
of the metrics of ``asymptotically AdS" Euclidean spaces. The
formal definition of such metrics is discussed at length in
\cite{kn:mcinnes}, and we need not rehearse all the details here:
the main point is simply as follows. Under conditions which will
always be satisfied for the spaces discussed here, the metric of
an asymptotically AdS Euclidean space M [with asymptotic sectional
curvature $-$1/L\2] can be written, near to any connected
component of the conformal boundary, as
\begin{equation}\label{eq:ARMADILLO}
g^+(\textup{M})\; =
\;{{\textup{L}^2}\over{\rho^2}}[d\rho^2\;+\;g^+_\rho],
\end{equation}
where $\rho$ is a coordinate such that the given component of the
conformal boundary is at $\rho$ = 0. Here $g^+_\rho$ is a metric
on the spaces transverse to the boundary. The point we wish to
stress is that $g^+_\rho$ \emph{does} in general have a
non-trivial dependence on $\rho$; the conformal structure at this
component of infinity is represented by a metric which is obtained
by taking the \emph{limit} of $g^+_\rho$ as $\rho$ tends to zero.
In this sense, the metrics of the form (\ref{eq:DRONGO})
considered above were very special cases, since we did not need to
take this limit. A good example of this limiting process is
provided by Lorentzian AdS\4 itself: in global coordinates
(t,r,$\theta,\phi$) the metric can be expressed as
\begin{eqnarray}\label{eq:DRONGOSON}
g^-(\mathrm{AdS}_4)\;= \; \mathrm{cosh}^2(\mathrm{r}/\mathrm{L})
\; [-\,\mathrm{dt}^2 \;+\;
\mathrm{sech}^2(\mathrm{r}/\mathrm{L})\; \mathrm{dr}^2
\;+\;\mathrm{L}^2\, \mathrm{tanh}^2(\mathrm{r}/\mathrm{L})\,
[d\theta^2 + \mathrm{sin}^2(\theta)d\phi^2]],
\end{eqnarray}
and we see that the metric still depends on r even after the
divergent conformal factor
$\mathrm{cosh}^2(\mathrm{r}/\mathrm{L})$ is removed. [Here of
course the boundary is obtained by letting r tend to infinity, so
that $\mathrm{tanh}^2(\mathrm{r}/\mathrm{L})$ tends to unity and
we obtain the usual cylindrical conformal boundary of AdS\4.]

This kind of behaviour is actually quite well-adapted to the
cosmological case, since it is well known [see for example
\cite{kn:turok1}] that the approach to cosmological singularities
is \emph{ultralocal}: that is, ultimately, only the [proper] time
dependence of the metric is important. Hence, in studying the very
late or very early stages of a Bang/Crunch cosmology, we can
indeed concentrate on metrics which resemble (\ref{eq:ARMADILLO}),
in the sense that the metric at infinity is obtained by stripping
away a conformal factor and then taking the limit of a family of
metrics parametrized by a single parameter. In the notation of
\cite{kn:turok1}, we can express the metric in the ultralocal
phase as
\begin{equation}\label{eq:DINO}
g^-_{\mathrm{Anisotropic}}\;=\;-\,(dt^-)^2\;+\;(a^-)^2\,\sum_{i}\mathrm{e}^{2\beta_i}\;(\sigma^i)^2,
\end{equation}
where $a^-(t^-)$ is an overall scale factor, where the $\sigma^i$
are orthogonal, time-independent one-forms on the spatial slices,
where the $\beta_i$ are three distinct functions of proper time
satisfying
\begin{equation}\label{eq:SAUR}
\beta_1\;+\;\beta_2\;+\;\beta_3\;=\;0,
\end{equation}
and where all dependence on spatial position has been suppressed.
For locally flat spatial sections one finds that
\begin{equation}\label{eq:STEGO}
{{\mathrm{d}\beta_i}\over{\mathrm{d}t^-}}\;=\;c_i\,(a^-)^{-3},
\end{equation}
and one can show \cite{kn:turok1} that the scale factor satisfies
a FRW equation of the form
\begin{equation}\label{eq:SAURUS}
3\,\mathrm{H}^2\;=\;8\pi\,[\rho\;+\;{{\sigma^2}\over{(a^-)^6}}],
\end{equation}
where H is the Hubble parameter, where $\rho$ is the total energy
density and where
\begin{equation}\label{eq:TYRANNOSAURUS}
\sigma^2\;=\;{{1}\over{2}}[c_1^2\;+c_2^2\;+c_3^2];
\end{equation}
thus $\sigma$ is a constant which is an overall measure of the
extent of anisotropy in such a spacetime.

In our case, $\rho$ is the sum of the energy density of the
background AdS\4, namely $-\,3/8\pi\,$L\2, with the energy density
of the quintessence field. Now with regard to this latter, recall
that we saw that the Breitenlohner-Freedman bound requires that
the series in equation (\ref{eq:ABRACADABRA2}) should terminate,
with the last value of k being the largest integer satisfying
\begin{equation}\label{eq:gamma}
{{2}\over{1\;+\;\sqrt{2}}}\;\times
{{1}\over{3}}\;\leq\;\varpi/\mathrm{k}^2\;.
\end{equation}
For example, in the case of $\varpi$ = 10 [see
\cite{kn:answering}], the last value of k is 6, and this means
that the corresponding quintessence component has a density
proportional to $(a^-)^{-\,7.2}$. [Recall that the magnitude of
the exponent must not exceed 7.2426.] In general, if the last
value of k satisfies (\ref{eq:gamma}), then it \emph{may} also
satisfy $\varpi/\mathrm{k}^2\,\ < \;1/3$. If this is so, then we
see from equation (\ref{eq:EXTRA}) that the quintessence energy
density grows, as $a^-$ tends to zero, more rapidly than
$(a^-)^{-\,6}$. For example, in the case where $\varpi$ = 10, this
means that, extremely near to the Bang or the Crunch ---
\emph{not} at other times --- equation (\ref{eq:SAURUS}) becomes
\begin{equation}\label{eq:STEGOSAURUS}
3\,\mathrm{H}^2\;=\;8\pi\,[\,{{-\,3}\over{8\pi\textup{L}^2}}\;+\;{{3}\over{8\pi\textup{L}^2\,(a^-)^{7.2}}}\;+\;{{\sigma^2}\over{(a^-)^6}}\,],
\end{equation}

Clearly, the second term on the right is the dominant one near to
the Bang and the Crunch --- and this would remain true even if we
included  the contributions of ordinary matter, radiation, and so
on. In particular, whatever the initial anisotropy $\sigma$ may
have been, it will be completely insignificant compared to this
term: one has a kind of ``cosmic no-hair" theorem.

The situation here is exactly analogous to the way, as one moves
\emph{away} from the initial singularity, the inflaton potential
dominates all other terms in the Friedmann equation, so that
anisotropies are ``inflated away" by the inflationary expansion:
here the ``last" quintessence component has the same effect as the
singularities are \emph{approached}, because its density grows
more rapidly than that of any other contribution. Since there is
no limit to the contraction, there is no limit to this effect ---
all local anisotropies will be completely wiped out in the very
last stages of the approach to the singularities. A very similar
phenomenon plays a crucial role in the cyclic cosmologies
\cite{kn:turok2}, and we see that it is equally important here,
although we stress that there is \emph{no} ``bounce" in our case:
we need rapid density growth rates not to prepare for a phase of
expansion succeeding a crunch, but to ensure that the metric
induced on [Euclidean] infinity is indeed flat. [Because of this
difference, it turns out that much larger values of the effective
equation-of-state parameter are required in the cyclic case than
here; as we know, in our case the magnitude of the largest
exponent of the scale factor is never much larger than six.]

The essential point here is that a three-dimensional Riemannian
manifold which is locally isotropic around each point --- that is,
there is a local isometry mapping any unit vector at any point to
any other unit vector at that point --- has a sectional curvature
which is independent of direction. For in \emph{three} dimensions
each unit vector at a point uniquely determines a two-dimensional
subspace of the tangent space, namely, the subspace perpendicular
to it. But if the sectional curvature of a Riemannian manifold of
dimension at least three is independent of direction, then
[\cite{kn:kobayashi}, page 202] it is also independent of
position; that is, the curvature is constant. Since compact
manifolds of constant negative curvature are in Kazdan-Warner
class [N], while those of constant positive curvature are in [P],
it follows that the only way that a metric on a manifold with the
topologies we are considering here can be locally isotropic is by
being perfectly flat. We conclude that the conformal structure
induced at Euclidean infinity is represented by a perfectly flat
metric, provided that the matter content of our spacetime is such
that all local anisotropies are eliminated by a ``final"
quintessence component with $\varpi/\mathrm{k}^2\,\ < \;1/3$.

We require, then, that the final value of k should satisfy
\begin{equation}\label{eq:delta}
{{2}\over{1\;+\;\sqrt{2}}}\;\times
{{1}\over{3}}\;\leq\;\varpi/\mathrm{k}^2\;<\;1/3.
\end{equation}
These inequalities express the competing demands of the
Breitenlohner-Freedman bound [which requires the lower bound] and
of Seiberg-Witten instability [which, via the
Schoen-Yau-Gromov-Lawson theorems, requires the upper bound]. It
is striking that the allowed interval is so short.

The effect of (\ref{eq:delta}) is to exclude certain values of
$\varpi$; the only allowed values are those lying in intervals [a
, b) where (\ref{eq:delta}) is satisfied for some integer k. These
intervals are given in the table. The intervals are closed to the
left, open to the right [so that, for example, $\varpi$ = 3 is
\emph{not} permitted].

\begin{center}
\begin{tabular}{|c|c|c|}
  \hline
k & a & b \\
\hline
1 &  0.276142 & 0.333333 \\
2 &  1.104570 & 1.333333 \\
3 &  2.485281 & 3.000000 \\
4 &  4.418278 & 5.333333 \\
5 &  6.903559 & 8.333333 \\
6 &  9.941126 & 12.000000 \\
7 &   13.530976 & 16.333333 \\
8 &   17.673112 & 21.333333 \\
9 &   22.367532 & 27.000000 \\
10 &   27.614238 & 33.333333 \\
11 &   33.413227 & 40.333333 \\
12 &   39.764502 & 48.000000 \\
  \hline

\end{tabular}
\end{center}
Notice that there is an upper bound on the values of $\varpi$ so
excluded, because the allowed interval for k = 11 overlaps the
allowed interval for k = 12, and all subsequent allowed intervals
overlap their successors. [One sees this either by consulting the
table or by means of a simple calculation based on requiring the
lower end of one interval to be smaller than the upper end of its
predecessor.] This upper bound is given by
\begin{equation}\label{eq:epsilon}
\varpi_{\mathrm{forbidden}}\; < \; {{242}\over{3}}\;(\sqrt{2} \; -
\; 1) \; \approx \; 33.4132.
\end{equation}
That is, all values of $\varpi$ above this number are allowed.
Below it, there is a haphazard set of intervals which are allowed,
alternating with intervals which are not. For example, $\varpi$ =
10, an example studied in detail in \cite{kn:answering}, is
allowed; on the other hand, $\varpi$ = 9.90 is not; nor is
$\varpi$ = 2, also studied in \cite{kn:answering}. In short,
values of $\varpi$ below 33.4132 entail careful fine-tuning;
larger values do not. If we can argue on independent grounds that
$\varpi$ is large, then there are no difficulties with fine
tuning.

We conclude that if we take $\varpi$ to be large, then ``cosmic
baldness" will automatically ensure that the conformal structure
induced at Euclidean infinity is represented by an exactly flat
metric, and this can be achieved without violating the
Breitenlohner-Freedman bound. In fact, the observations
\cite{kn:answering} require $\varpi$ to be at least this large.
Furthermore, there are general theoretical reasons for expecting
$\varpi$ to be larger still. In many string theory
compactifications \cite{kn:denef}\cite{kn:bobby} there is a
general tendency to predict that the fundamental length scale of
our observed spacetime should be very short. But one sees from
equations (\ref{eq:M}) and (\ref{eq:L}) that the natural length
scales of the Euclidean and Lorentzian versions of our spacetime
are different: in the former case, the space is asymptotic to a
Euclidean AdS\4 with ``radius" L, whereas in the latter case the
Universe is finite in all directions, including time, with a total
lifetime of $\pi\varpi$L. Thus L can indeed be small without
contradicting the observations, provided that $\varpi$ is very
large.

To summarize: Seiberg-Witten instability allows us, in the context
of cosmic holography, to draw several surprisingly strong
conclusions regarding the spatial sections of the Universe. The
first is that negatively curved compact spatial sections are
completely ruled out in string theory. In this case, the
instability is particularly persistent, because it is
\emph{topological}: the spatial sections can be arbitrarily
deformed as we trace them back to the Bang or forward to the
Crunch, yet the system is still subject to instabilities arising
from the nucleation of branes which lower their action as they are
moved towards the [Euclidean] boundary.

Combining this with Linde's \cite{kn:lindetypical} analysis
discussed earlier, we find that \emph{the flat compact
three-manifolds are unique} in their ability to satisfy all of the
strictures imposed by the requirements of Inflation [which
accommodates positively curved sections only with great
difficulty] and large brane instability [which even more firmly
rules out negatively curved sections]. Even the flat manifolds
only narrowly escape Seiberg-Witten instability. They escape it if
we can make the boundary \emph{exactly} conformally flat; we saw
that our ``toy model" of an accelerating holographic cosmos was
able to perform this feat, provided that the parameter $\varpi$ is
sufficiently large, as is naturally the case.

The conclusions we have reached here, while developed in the
context of a particular model, are in fact extremely robust. That
is, they do not depend on the particular choice we made --- a
Euclidean axion --- for the matter content of our cosmology: for
example, the prohibition on negatively curved spatial sections is
extremely general, since it depends only on the topology of these
spaces and not on their geometry. Our ability to avoid
Seiberg-Witten instability in the case of a boundary in KW class
[Z] did depend on the ability of our matter model to flatten the
sections as the singularities are approached, but this is
attainable for many matter models --- see \cite{kn:turok2}.

However, we shall now see that the list of candidates for the
spatial geometry of the world can be still further reduced if we
do adopt the specific matter model introduced earlier. Thus the
findings of the next section should not be considered to be as
general as those of this section.

\addtocounter{section}{1}
\section*{\large{5. Finite --- and [Conformally] Flat --- In All Directions}}
With the help of Seiberg-Witten instability \cite{kn:seiberg},
cosmic holography \cite{kn:maoz}, and Inflationary arguments
\cite{kn:lindetypical}, we have reduced the number of candidates
for the spatial sections of the Universe to a mere ten. That is
nevertheless nine too many.

If we knew precisely which of these ten has been chosen by Nature,
then we would have a valuable clue as to the true nature of the
initial state. The ten candidates, dubbed the \emph{platycosms} by
Rossetti and Conway \cite{kn:conway} [a term we adopt here as a
useful abbreviation], are of varying degrees of complexity. Among
those which are orientable, we have the torus T\3, the dicosm
T\3/$\bbz_2$, the tricosm T\3/$\bbz_3$, the tetracosm
T\3/$\bbz_4$, the hexacosm T\3/$\bbz_6$, and the didicosm or
Hantzsche-Wendt space T$^3$/[$\bbz_2\;\times\;\bbz_2$].

For all that we know, the spatial sections of our Universe could
have the structure of the didicosm. Unlike the torus, this space
has non-trivial holonomies, of two different kinds: the holonomy
group is $\bbz_2\;\times\;\bbz_2$, a finite subgroup of SO(3). The
fundamental domain here is a rhombic dodecahedron, and, if this
domain were small enough to be observable, the resulting patterns
in the microwave sky would be remarkable indeed \cite{kn:weeks}.
Even if it is not directly observable, a theoretical deduction
that the spatial sections have such a  complicated structure would
surely be a strong hint that the initial state has been selected
with great precision, presumably by something very much more
intricate than a simple classical singularity. But how can such a
theoretical deduction be made? In this section, we shall show how
our toy model, with Euclidean metric (\ref{eq:M}) [where the
transverse sections are not necessarily globally T\3] leads to a
partial answer to this question. The hope of course is that a more
realistic matter model would yield a more complete answer.

Equation (\ref{eq:M}) indicates that \emph{if} we interpret the
underlying manifold as the interior of a manifold-with-boundary,
then that boundary is disconnected: it has two connected
components. It was emphasised by Maldacena and Maoz \cite{kn:maoz}
themselves that the status of Euclidean manifolds with a
disconnected boundary is very problematic from a holographic point
of view. In general this apparent failure of a one-to-one
correspondence is a very deep question [see
\cite{kn:yau}\cite{kn:balasub}\cite{kn:gukov} for discussions],
but in \cite{kn:answering} we suggested that it may have a very
simple resolution \emph{in the particular case} with which we are
concerned here. The argument is as follows.

The two-dimensional open cylinder (0, 1) $\times$ S$^1$ can be
compactified in [at least] two different ways. The first is to
regard it as the interior of the compact manifold-with-boundary
[0, 1] $\times$ S$^1$ [the closed cylinder]; the second is to
regard it as an open submanifold of the torus S$^1$ $\times$ S$^1$
= T\2 [obtained from T\2 by deleting a circle]. Neither option is
``correct": one makes a choice depending on the circumstances. The
difference, of course, is that in the first case we have to add
\emph{two} circles, whereas in the second case we only need to add
one. This led us, in \cite{kn:answering}, to suggest that the
second kind of interpretation is required by cosmic holography.

In the case at hand, we can re-express the metric (\ref{eq:M}) in
the following way. Define a constant c$_{\varpi}$ by
\begin{equation}\label{eq:R}
\textup{c}_{\varpi}\; =
\;{{\varpi}\over{\pi}}\int_{0}^{\infty}\textup{sech}^{\varpi}(\zeta)d\zeta
\;,
\end{equation}
and a new coordinate $\theta$ by
\begin{equation}\label{eq:RATS}
\textup{c}_{\varpi}\textup{Ld}\theta \;=
\;\pm\textup{sech}^{\varpi}({{t^+}\over {\varpi \textup{L}}})dt^+,
\end{equation}
where the sign is + when t$^+$ is positive, $-$ when t$^+$ is
negative. The range of $\theta$ is $-\pi$ to $+\pi$. Now solve for
t$^+$ in terms of $\theta$ and use this to express
sech$^{\varpi}$(${{\textup{t}^+}\over {\varpi \textup{L}}}$) in
terms of $\theta$. Denote this function by G$_{\varpi}(\theta)$;
then G$_{\varpi}(\theta)$ vanishes at $\pm\pi$, and
$g^+(\varpi,A)$ is given in terms of the coordinate $\theta$ as
\begin{equation}\label{eq:S}
g^+(\varpi,A) \;=\;
\textup{c}^2_{\varpi}\textup{L}^2\;\textup{G}^{-2}_{\varpi}(\theta)\;[\,d\theta^2
\; +\;
({{A}\over{\textup{c}_{\varpi}\,\textup{L}}})^2\,(d\theta_1^2
\;+\; d\theta_2^2 \;+\; d\theta_3^2)].
\end{equation}
As it stands, the coordinate $\theta$ cannot be extended to the
whole circle: we have to delete the [\emph{single}] point $\theta$
= $\pm\pi$, because $g^+(\varpi,A)$ is singular there. However,
removing the prefactor on the right side of (\ref{eq:S}) by a
conformal transformation, we obtain precisely the standard local
metric for a non-cubic four-dimensional torus. [The number
A/c$_{\varpi}$L can in fact be constrained by observational data:
see \cite{kn:answering}.]

If we had begun with (\ref{eq:S}) instead of (\ref{eq:M}), we
would undoubtedly have declared that the natural compactification
of our Euclidean space is a space with the local geometry of a
torus. Infinity here is not a boundary; it is instead a
``submanifold at infinity". The real point, however, is that
infinity is \emph{connected} in this interpretation. Clearly the
``double boundary" problem simply does not arise if we adopt this
viewpoint.

Thus, we propose an extremely simple extension of our hypothesis
of flat, compact spatial sections: not only the spatial sections,
but also [the compactified Euclidean version of] \emph{the entire
four-dimensional spacetime} should have the topology of a compact
flat manifold. In short, the spatial sections are flat, compact
three-manifolds, while the compactified spacetime is globally
conformal [equation (\ref{eq:S})] to a flat, compact
four-manifold.

Now recall that the adoption of the local three-dimensional metric
$A^2\;[d\theta_1^2\; + \;d\theta_2^2 \;+\; d\theta_3^2]$ in
equation (\ref{eq:I}) did not commit us to the global geometry and
topology of the three-dimensional torus: \emph{many} distinct
compact flat three-dimensional manifold have this local metric,
since all such manifolds can be expressed topologically as T\3/F,
for some finite group F. [Recall that we assumed for simplicity
that the covering torus was cubic, but trivial modifications allow
us to consider the most general case.] In the same way, the
appearance of the metric of a four-dimensional torus in
(\ref{eq:S}) does not mean that we have here a manifold with the
topology of T$^4$ or even that of S$^1\;\times$ T\3/F. In
constructing a general compact flat manifold, the procedure is as
in the familiar case of a torus, but one is free to apply various
isometries before performing the identifications that produce a
compact space. In three dimensions, there are ten ways of doing
this; in four dimensions \cite{kn:wolf}, there are no fewer than
75, though we hasten to add that most of these 75 cannot be
constructed from manifolds of the form T\3/F in the above way.

Return temporarily to the interpretation of $g^+(\varpi,A)$ as a
metric on the interior of a manifold-with-boundary. That boundary
consists of two copies of T\3/F. What we are proposing here is
that these two copies should be identified. But, before performing
the identification, we are free to apply an isometry, as above, to
one of the copies. If we do this, we shall obtain a space with the
same local metric as S$^1\;\times$ T\3/F [which is what we get if
the isometry is trivial]; that is, we obtain a metric of the form
(\ref{eq:S}). The spaces we obtain will not be fully general flat
compact four-dimensional manifolds, partly because the metric in
(\ref{eq:S}) has a special form [the torus is rectangular and has
three dimensions of the same length, while the fourth in general
has a different length] and partly because we have already fixed
part of the topology by specifying the group F. Nevertheless some
freedom remains, because T\3/F still has a non-trivial isometry
group even after the factoring by F. Can we reduce this freedom in
a physical way? The answer is that we can, by exploiting the fact
that our matter field is of a very specific kind: it is a
[Euclidean] \emph{axion}.

The axion field $\varphi^+$ is able to distinguish t$^+$ =
$-\;\infty$ from t$^+$ = $+\;\infty$, because from equation
(\ref{eq:K}) we have
\begin{equation}\label{eq:STAR}
\lim_{t^+\to\infty}\;\varphi^+\;=\;
\sqrt{{{\varpi}\over{4\pi}}}\times{{\pi}\over{2}}\;=\;-\;\lim_{t^+\to-\;\infty}\;\varphi^+,
\end{equation}
and this sign difference is physical because from equation
(\ref{eq:C}) we have
\begin{equation}\label{eq:V}
\textup{W}^+(-\sqrt{{{\varpi}\over{4\pi}}}\times{{\pi}\over{2}})\;=\;-\;
\textup{W}^+(+\sqrt{{{\varpi}\over{4\pi}}}\times{{\pi}\over{2}}).
\end{equation}
This is important, because it apparently puts a stop to our plan
of identifying the two boundary components: how can we do so when
the field and its superpotential take different values on each
component? But recall that an axion naturally reverses sign under
a reversal of orientation. Thus the problem is solved in a
natural, geometric way if --- and only if --- we arrange for the
orientation of one boundary component to be reversed before
identifying it with the other. This is precisely the way, in two
dimensions, a Klein bottle \cite{kn:wolf} is defined, and we can
proceed in much the same way here, allowing however for the
greater complexity of T\3/F. Let us see to what extent this
requirement reduces our freedom in constructing our
four-dimensional Euclidean ``spacetime".

The point is simply that not every compact flat three-manifold
admits an orientation-reversing isometry. In essence, factoring a
Riemannian manifold by a discrete group usually reduces the
``size" of the isometry group, since not all isometries of the
original space are compatible with the factoring. In the present
case, the reduction can obstruct the procedure we outlined above.
The isometry groups of all of the platycosms are listed in
\cite{kn:conway}, and the result we need can simply be stated
using that list; however, we can gain more insight by means of the
following elementary argument.

Suppose that one has a manifold M admitting a group G(M) of
diffeomorphisms (such as isometries, conformal symmetries, and so
on). Let $\Gamma$ be a subgroup of G(M) and let N($\Gamma$) be the
\emph{normalizer} of $\Gamma$ in G(M). That is,
\begin{equation}\label{eq:eighteen}
\mathrm{N}(\Gamma) = \{g \in \mathrm{G(M)} \;\mid \; g\gamma
g^{-1} \in \Gamma \;\;\forall \;\;\gamma \in \Gamma\}.
\end{equation}
Clearly N($\Gamma$) contains all those elements of G(M) which
descend to well-defined diffeomorphisms of M/$\Gamma$. But notice
that every element of $\Gamma$ itself has no effect on each
element of M/$\Gamma$. Thus the symmetry group of M/$\Gamma$,
which we denote by G(M/$\Gamma$), is not N($\Gamma$) but rather
the quotient N($\Gamma$)/$\Gamma$:
\begin{equation}\label{eq:twenty}
\mathrm{G(M}/\Gamma) = \mathrm{N}(\Gamma)/\Gamma.
\end{equation}
See \cite{kn:orbifold} for more details and for other applications
of this formula.

Let us see how this works in some concrete examples. First, the
torus T\3 is defined as follows [\cite{kn:wolf}, page 117]. First
recall that any isometry of $\bbr^3$ can be expressed as (B, a),
where B is an orthogonal matrix and a is a vector, and where (B,
a) means that we let B act first, followed by a translation
through a. Let a$_i$, i = 1,2,3 be a fixed basis for $\bbr^3$. If
$\Gamma_3^*$ is generated over the integers by the isometries
$\tau_i$ = (I$_3$, a$_i$), where I$_3$ is the identity matrix,
then T\3 = $\bbr^3/\Gamma_3^*$. Now consider the isometry $\Omega$
= ($-\,\mathrm{I}_3$, 0). It is easy to see that conjugation by
$\Omega$ just maps each element of $\Gamma_3^*$ to its inverse.
Thus $\Omega$ does normalize $\Gamma_3^*$ and so it projects to an
isometry of T\3. Of course, $\Omega$ reverses the orientation of
$\bbr^3$, so we see that T\3 does admit an orientation-reversing
isometry, which is what we need.

Next, the platycosm of the form T\3/$\bbz_3$ [the tricosm] is
obtained as follows. First we constrain the vectors a$_i$: we
require a$_1$ to be orthogonal to the other two, and we require
a$_2$ and a$_3$ to be of the same length and to be inclined at an
angle of 2$\pi$/3. Next, we set $\alpha$[tri] = (A$_3$[2$\pi$/3],
a$_1$/3), where A$_3$[2$\pi$/3] is a 3 $\times$ 3 matrix
corresponding to a rotation through 2$\pi$/3 in the a$_2$, a$_3$
plane: that is, A$_3$[2$\pi$/3] maps a$_1$ to itself, a$_2$ to
a$_3$, and a$_3$ to $-$a$_2$ $-$ a$_3$. Then the tricosm is
$\bbr^3$/$\Gamma$[tri], where $\Gamma$[tri] is obtained by
adjoining $\alpha$[tri] to the generators of $\Gamma_3^*$. Now
conjugation by $\Omega$ still maps each element of $\Gamma_3^*$ to
its inverse, but it does not have this effect on $\alpha$[tri];
instead we have
\begin{equation}\label{eq:KENYA}
\Omega\,\alpha[\mathrm{tri}]\,\Omega^{-1}\;=\;(\mathrm{A}_3[2\pi/3],
- \,\mathrm{a}_1/3).
\end{equation}
The isometry on the right is not the inverse of $\alpha$[tri] and
is \emph{not} an element of $\Gamma$[tri]. Thus $\Omega$ does not
descend to an isometry of the tricosm. In fact, in order to do
this, an isometry of $\bbr^3$ would have to reverse orientation in
the plane defined by a$_2$ and a$_3$, while also reversing a$_1$;
but such an isometry could not be orientation-reversing.

A similar argument works also for the tetracosm T\3/$\bbz_4$ and
the hexacosm T\3/$\bbz_6$. It does not work for the dicosm
T\3/$\bbz_2$. To see why, note that this space is defined much as
the tricosm, except that apart from being orthogonal to a$_1$, the
other conditions on a$_2$ and a$_3$ are dropped, and $\alpha$[tri]
is replaced by $\alpha$[di] = (A$_3$[$\pi$], a$_1$/2), where
A$_3$[$\pi$] is a 3 $\times$ 3 matrix rotating the a$_2$, a$_3$
plane through $\pi$; then the dicosm is $\bbr^3$/$\Gamma$[di],
where $\Gamma$[di] is obtained by adding $\alpha$[di] to the
generators of $\Gamma_3^*$. Again, conjugation by $\Omega$ maps
$\Gamma_3^*$ to itself, but now we have
\begin{equation}\label{eq:UGANDA}
\Omega\,\alpha[\mathrm{di}]\,\Omega^{-1}\;=\;(\mathrm{A}_3[\pi], -
\,\mathrm{a}_1/2),
\end{equation}
and this \emph{is} in $\Gamma$[di] since it is the inverse of
$\alpha$[di] [because A$_3$[$\pi$] is of order two]. Thus $\Omega$
does descend to an orientation-reversing isometry of the dicosm.
The didicosm T$^3$/[$\bbz_2\;\times\;\bbz_2$] can be constructed
in much the same way as the dicosm, but with three additional
generators instead of one, each involving a rotation by $\pi$ in
some plane. One can show that $\Omega$ descends to an
orientation-reversing isometry in this case also. In the case of
the non-orientable platycosms, a different argument applies, but
for those platycosms it is in any case obvious that there can be
no orientation-reversing isometries. \emph{Thus the torus, the
dicosm, and the didicosm are the only survivors}, that is, they
are the only platycosms which can be used to construct a
``generalized Klein bottle" of the kind we need in order to
compensate for the fact that the axion field has opposite signs on
the two connected components of infinity in our model.

Let us now show how to construct the final compact flat
four-manifolds which, as explained above, are the possible
underlying spaces of the conformal compactification corresponding
to the metric (\ref{eq:S}). We shall concentrate on the dicosm:
the construction for the torus is just a simpler version, while
that for the didicosm is somewhat more complicated but introduces
no essentially new difficulties.

Let a$_{\mu}$, where $\mu$ = 0 through 3, be an orthogonal basis
for $\bbr^4$, where we take a$_0$ to be of length
2$\pi$c$_{\varpi}$L, while a$_1$, a$_2$, and a$_3$ are of length
2$\pi$A. Define 4 $\times$ 4 matrices A$_4$ and B$_4$ by A$_4$ =
diag(1, 1, $-$1, $-$1) and B$_4$ = diag(1, $-$1, $-$1, $-$1), and
then define a pair of $\bbr^4$ isometries as follows.
\begin{equation}\label{eq:ZAIRE}
\alpha \;=\;(\mathrm{A}_4,\;\mathrm{a}_1/2),
\;\;\beta\;=\;(\mathrm{B}_4,\;\mathrm{a}_0/2).
\end{equation}
Let $\Delta_4(\Gamma[\mathrm{di}])$ be the group generated by
these isometries, together with the translations $\tau_\mu$ =
(I$_4$, a$_\mu$). Then $\Delta_4(\Gamma[\mathrm{di}])$ can be
presented as follows:
$$
\begin{array}{rclrclrcl}\label{eq:FINARFIN}
 \nonumber\alpha^2 &  = & \tau_1, &  \beta^2 & = &  \tau_0,
 & \beta\alpha\beta^{-1} & = & \alpha^{-1},  \\
 \nonumber\alpha\tau_0\alpha^{-1} & =  &
 \tau_0, & \alpha\tau_2\alpha^{-1}
 & = & \tau_2^{-1}, & \alpha\tau_3\alpha^{-1} & = &\tau_3^{-1},
  \\
\beta\tau_1\beta^{-1} & = & \tau_1^{-1}, & \beta\tau_2\beta^{-1} &
= & \tau_2^{-1}, & \beta\tau_3\beta^{-1} & = & \tau_3^{-1}.
\end{array} $$
$\Delta_4(\Gamma[\mathrm{di}])$ is so named because it contains a
subgroup $\Gamma[\mathrm{di}]$, generated by $\alpha$, $\tau_1$,
$\tau_2$, and $\tau_3$, which corresponds to the fundamental group
of the dicosm. One can see that $\Delta_4(\Gamma[\mathrm{di}])$ is
a non-abelian infinite group with no element of finite order
[other than the identity] and with a maximal free abelian subgroup
$\Gamma_4^*$ [generated by the $\tau_{\mu}$] consisting of four
copies of $\bbz$. From the relations given, it is clear that
$\Gamma_4^*$ is normal in $\Delta_4(\Gamma[\mathrm{di}])$; the
quotient $\Delta_4(\Gamma[\mathrm{di}])$/$\Gamma_4^*$ is of finite
order [it is isomorphic to $\bbz_2\;\times\;\bbz_2$]; one says
that $\Gamma_4^*$ is of index 4 in $\Delta_4(\Gamma[\mathrm{di}])$
and of rank 4. By the relevant version of the Bieberbach theorems
[\cite{kn:wolf}, page 105] it follows that
$\bbr^4/\Delta_4(\Gamma[\mathrm{di}])$ is a four-dimensional
manifold, covered by a four-torus $\bbr^4/\Gamma_4^*$ with the
flat metric
\begin{equation}\label{eq:WHATEVER}
g_{\mathrm{flat}}^+(\mathrm{c_{\varpi}L},\,\mathrm{A}) \;=\;
\mathrm{c_{\varpi}^2L^2}\,d\theta^2 \; +\;
\mathrm{A}^2\,(d\theta_1^2 \;+\; d\theta_2^2 \;+\; d\theta_3^2);
\end{equation}
here $\theta$ is an angular coordinate corresponding to a$_0$,
while the $\theta_i$ correspond to the a$_i$.

As a Riemannian manifold, $\bbr^4/\Delta_4(\Gamma[\mathrm{di}])$
can be expressed  as T$^4$/[$\bbz_2\;\times\;\bbz_2$], where T$^4$
is the rectangular torus with aspect ratio given by
A/(c$_{\varpi}$L), and where $\bbz_2\;\times\;\bbz_2$ is the
linear holonomy group of this space. One of the two independent
non-trivial holonomies reverses orientation, while the other does
not. All this can be repeated beginning with the three-torus
instead of the dicosm, resulting in a flat four-manifold with the
structure T$^4$/$\bbz_2$, or with the didicosm [Hantzsche-Wendt
space], resulting in a flat four-manifold with the structure
T$^4$/[$\bbz_2\;\times\;\bbz_2\;\times\;\bbz_2$].

The overall picture, then, is this. The underlying structure of
the compactified Euclidean four-dimensional space is that of
T$^4$/$\bbz_2$, T$^4$/[$\bbz_2\;\times\;\bbz_2$], or
T$^4$/[$\bbz_2\;\times\;\bbz_2\;\times\;\bbz_2$]. Let us take this
last space as a concrete example. The local metric is given in
(\ref{eq:WHATEVER}); it is indistinguishable from that of a
four-torus, except that the coordinates are not global. However,
if we move around the ``time" direction [the $\theta$ direction],
we find that orientation is reversed once per cycle. If we
arbitrarily select $\theta$ = $\pm\pi$ to be the the [single]
point on the $\theta$-circle where orientation is reversed, then a
pseudoscalar such as our Euclidean axion will automatically
reverse sign there. Now suppose that we single out $\theta$ =
$\pm\pi$ by performing the conformal transformation that maps the
metric in (\ref{eq:WHATEVER}) to the one given in equation
(\ref{eq:S}). Now $\theta$ = $\pm\pi$ has to be excised, and the
space has the topology of a Bang/Crunch spacetime with spatial
sections having the structure of a didicosm. Transforming to the
Lorentzian version, we have the Bang/Crunch cosmology with metric
(\ref{eq:L}). In the reverse direction, we transform the metric in
(\ref{eq:L}) to its Euclidean version, which is apparently a space
with a conformal infinity consisting of two components, each
having the structure of the didicosm; however, we can naturally
identify these after performing an orientation-reversing twist, so
as to obtain the compact four-manifold
T$^4$/[$\bbz_2\;\times\;\bbz_2\;\times\;\bbz_2$] as the
compactification. The dual field theory resides on the
\emph{single}, orientable, didicosm section at $\theta$ =
$\pm\pi$, that is, at infinity.

Linde's considerations of quantum gravity \cite{kn:lindetypical},
with which we began, allowed the spatial sections of our universe
to have any one of the infinite variety of structures possible for
compact three-manifolds of negative or zero curvature. We have
narrowed this vast array down to just three candidates: the torus
T\3, the dicosm T\3/$\bbz_2$, and the didicosm
T\3/[$\bbz_2\;\times\;\bbz_2$]. We do not know how to reduce this
list to a single candidate. It is noteworthy, however, that
although they seem rather similar, the \emph{homology} groups of
the three surviving candidates are very different: in particular,
their first homology groups [with integer coefficients] are given
on page 122 of \cite{kn:wolf} as
\begin{eqnarray}\label{eq:HOMO}
\mathrm{H}_1[\mathrm{T}^3] & = & \bbz\;\times\;\bbz\;\times\;\bbz \nonumber\\
\nonumber\mathrm{H}_1[\mathrm{T}^3/\bbz_2] & = & \bbz\;\times\;\bbz_2\;\times\;\bbz_2\\
\mathrm{H}_1[\mathrm{T}^3/[\bbz_2\;\times\;\bbz_2]] & = &
\bbz_4\;\times\;\bbz_4.
\end{eqnarray}
Notice that this last group is \emph{finite}; thus the cycles
around which branes may be wrapped have a very different structure
in the didicosm from those in the torus or the dicosm. It also
follows that the first [and therefore second] Betti numbers are
quite different:
\begin{equation}\label{eq:BETTI}
\mathrm{b}_1[\mathrm{T}^3]\;=\;3,\;\;\;\;\;\;\mathrm{b}_1[\mathrm{T}^3/\bbz_2]\;=\;1,\;\;\;\;\;\;
\mathrm{b}_1[\mathrm{T}^3/[\bbz_2\;\times\;\bbz_2]]\;=\;0.
\end{equation}
The vanishing of
$\mathrm{b}_1[\mathrm{T}^3/[\bbz_2\;\times\;\bbz_2]]$ means that,
in sharp contrast to the torus and the dicosm, the didicosm has no
harmonic one-forms or two-forms. A further study of all of these
properties may lead to a physical way of distinguishing the
didicosm from all other candidates.

 \addtocounter{section}{1}
\section*{6. Conclusion}
What is the shape of space? This question, even in its modern
form, has exercised leading minds for over a century
[\cite{kn:schwarzschild}; see \cite{kn:mcinnesschwarz} for a
discussion of the views of de Sitter and Schwarzschild]. For a
time there was hope that it would be settled by direct
observation, but, as this hope has begun to fade, we may have to
turn to theory for guidance. There is a very large literature on
observational aspects of topologically non-trivial cosmological
models, but very little is known about the basic physical
principles which might prefer one topology over another. Indeed,
one of the main motivations of this work is to persuade the reader
that it is possible to find such principles.

It is striking and exciting that Inflation, which tells us that we
should probably not hope to see direct evidence of non-trivial
spatial topology, nevertheless also tells us
\cite{kn:lindetypical} that this topology probably \emph{is}
non-trivial. It then becomes a pressing question to determine
which topological structure has been chosen --- and how.

In this work, we have argued that our best theories of fundamental
physics do allow us to narrow the field of candidates. The lessons
we have learned vary in their degree of generality.

The most general lessons are based on the assumption that
\emph{some kind of bulk-boundary duality is valid in cosmology}.
This very general assumption already has strong consequences. Most
importantly, it tells us that we cannot ignore the most distant
regions of our world, those beyond cosmological horizons: for
those regions are just as surely part of the bulk as the regions
near to us in space and time, and their role in the boundary dual
theory cannot be excluded or neglected.

Slightly less generally, if we assume that the spacetime conformal
boundary lies to the future and the past [as in de Sitter
spacetime, or in any of the cosmologies of the general form
considered by Maldacena and Maoz \cite{kn:maoz}], then typically
each connected component of the ``dual" space has the same
topology as the spatial sections. But if we further assume that
string theory controls the bulk-boundary relationship, then the
very general arguments of Seiberg and Witten \cite{kn:seiberg}
apply. We are still at a very high level of generality at this
point, but already we can, with the help of the theorem of
Kazdan-Warner, make an extremely strong deduction about the nature
of the spatial sections: \emph{they cannot be negatively curved}.
The startling feature of this argument is precisely the fact that
it is \emph{topological}: once it has been established that a
manifold is in KW class [N], its scalar curvature must remain
negative [ensuring Seiberg-Witten instability] no matter how it
may be deformed by the subsequent evolution of spacetime.

If Seiberg-Witten instability rules out negative curvature, and
some of the most interesting versions of Inflation disfavour
positive curvature \cite{kn:lindetypical}, then of course we are
directed towards the flat, compact three-manifolds: the
\emph{platycosms} \cite{kn:weeks}\cite{kn:conway}. Already this is
a great reduction, since there are infinitely many compact
negatively curved spaces, but only ten platycosms.

All of these conclusions are very general, since they do not
depend on using a specific cosmological model. If we are willing
to be more specific, then we can reduce the list still further.
The particular cosmological model considered here, combined with
the [holographically motivated] requirement that disconnected
boundaries must be avoided, leads to a demand that the spatial
sections should have a specific geometric property; and we found
that only \emph{three} candidates satisfy this condition. We do
not ask the reader to take this particular conclusion very
seriously, since it is based on specific properties of a specific
matter model. The more important conclusion is that \emph{we have
shown explicitly that it is indeed possible to use physical
principles to effect a vast reduction in the range of candidates}.
It does not seem too far-fetched to imagine that a more
sophisticated and realistic matter model might well succeed in
reducing the list to a \emph{single} candidate. It may be that
this is how the shape of space will be discovered: perhaps only
one topology is consistent with our best theories.

Recent work on the [surprisingly deep] geometry of the platycosms
sheds interesting light on this observational/theoretical
interplay for cosmic topology. It has been found \cite{kn:doyle}
that it is possible for two platycosms with different topologies
to be \emph{isospectral}, that is, the spectra of their Laplace
operators can be placed into a one-to-one correspondence. This is
remarkable, because two three-dimensional tori can be isospectral
only if they have the same shape and size. Since the analysis of
CMB data involves precisely these spectra, it could be very
difficult to distinguish these two spaces by means of CMB
observations, even if the fundamental domain were small enough for
direct observations to be possible. Yet they are certainly
distinguished by our theoretical analysis above. For the
isospectral pairs are obtained by taking a certain specific
 torus T$_0^3$ of a fixed shape [it is a ``two-storey" rectangular
torus], and then taking quotients. The quotient of the form
T$_0^3$/[$\bbz_2\;\times\;\bbz_2$] is a particular example of a
didicosm, named ``Didi" in \cite{kn:doyle}; the quotient of the
form T$_0^3$/$\bbz_4$ is an example of a tetracosm, named
``Tetra"\footnote{The name Dexter has also been suggested.}. Didi
and Tetra are isospectral, but we saw above that Didi is
acceptable in our specific cosmology while Tetra must be excluded.
Thus we have a situation where theoretical arguments are able to
distinguish candidates which may be difficult to separate
observationally.

Of course the main task now is to determine or at least constrain
the boundary field theory. Because of the special asymptotic
properties of the Maldacena-Maoz cosmologies in the accelerating
case \cite{kn:mcinnes}, there is reason to believe that the
bulk/boundary correspondence will be unusual here. This is
currently under investigation.

 \addtocounter{section}{1}
\section*{Acknowledgements}
The author is extremely grateful to Wanmei for many encouraging
words.

\end{document}